\documentclass{IEEEtran}
\usepackage{cite}
\usepackage{amsmath,amssymb,amsfonts}
\usepackage{graphicx}
\usepackage{textcomp}
\usepackage{multirow}
\usepackage{xcolor}
\usepackage{eurosym}
\usepackage{fancyhdr}
\usepackage{algorithm}
\usepackage{comment}
\usepackage{bm}
\usepackage[noend]{algpseudocode}
\def\BibTeX{{\rm B\kern-.05em{\sc i\kern-.025em b}\kern-.08em
    T\kern-.1667em\lower.7ex\hbox{E}\kern-.125emX}}
\usepackage{amsmath}

\usepackage{eurosym}
\usepackage{nomencl}
\usepackage{multicol}

\ifCLASSOPTIONcompsoc
    \usepackage[caption=false, font=normalsize, labelfont=sf, textfont=sf]{subfig}
\else
\usepackage[caption=false, font=footnotesize]{subfig}
\fi

\newcommand{\N}{\mathcal{N}}

\begin{document}

\title{Arbitrage Tactics in Local Markets via Hierarchical Multi-agent Reinforcement Learning}

\author{Haoyang~Zhang,~\IEEEmembership{Graduate Student Member,~IEEE,} 
        Mina~Montazeri,~\IEEEmembership{Member,~IEEE,}
        Philipp~Heer,~\IEEEmembership{Member,~IEEE,}
        Koen~Kok,~\IEEEmembership{Senior Member,~IEEE,}
        and Nikolaos~G.~Paterakis,~\IEEEmembership{Senior Member,~IEEE}% <-this % stops a space

\thanks{H. Zhang, K. Kok, and N. G. Paterakis are with the Department of Electrical Engineering, Eindhoven University of Technology (TU/e), 5600 MB Eindhoven, The Netherlands (e-mail: h.zhang@tue.nl).}% 
\thanks{H. Zhang, M. Montazeri, and P. Heer are with the Urban Energy Systems Laboratory, Swiss Federal Laboratories for Materials Science and Technology (Empa), 8600 Dübendorf, Switzerland.}}%

\maketitle

\begin{abstract}
Strategic bidding tactics employed by prosumers in local markets, including the Local Electricity Market (LEM) and Local Flexibility Market (LFM), have attracted significant attention due to their potential to enhance economic benefits for market participants through optimized energy management and bidding. While existing research has explored strategic bidding in a single market with multi-agent reinforcement learning (MARL) algorithms, arbitrage opportunities across local markets remain unexplored. This paper introduces a hierarchical MARL (HMARL) algorithm designed to enable aggregator arbitrage across multiple local markets. The strategic behavior of these aggregators in local markets is modeled as a two-stage Markov game: the first stage involves the LEM, while the second stage encompasses both the LFM and the balancing market. To solve this two-stage Markov game, the HMARL framework assigns two sub-agents to each aggregator, a primary sub-agent and a secondary sub-agent. Without the arbitrage strategy, these sub-agents operate in silos, with the primary sub-agent focusing on first-stage profits and the secondary sub-agent on second-stage profits, each employing independent MARLs. On the contrary, when implementing the arbitrage strategy with the proposed HMARL, the sub-agents communicate and coordinate to perform arbitrage across multiple local markets, enhancing overall efficiency. The case study, conducted under a scenario where all aggregators employ the arbitrage strategy, shows that despite higher initial costs in the LEM, this strategy generates substantial savings in the LFM and the balancing market, resulting in a total profit increase of $40.6\%$ on average. This highlights the capability of the proposed HMARL to address the two-stage Markov game and facilitate arbitrage across local markets, thereby enhancing profitability for participants.\looseness=-1
\end{abstract}


\section{Introduction}

The emergence of Local Electricity Markets (LEM) and Local Flexibility Markets (LFM) within distribution networks (DNs) has fundamentally shifted energy trading and management towards a consumer-centric paradigm. These local markets facilitate decentralized energy systems by optimizing resource allocation and enhancing grid stability through localized and efficient trading mechanisms. Recent studies have proposed hierarchical local market structures comprising LEM and LFM \cite{hierarchical1, hierarchical2, hierarchical3}. Within these hierarchical local markets, prosumers with energy resources and loads initially participate in an LEM to trade electricity. Following the clearing of an LEM and the establishment of scheduled profiles, an LFM is introduced to enable aggregators to provide flexibility services, which are crucial for ensuring the safe operation of the associated DN, particularly in the presence of intermittent renewable energy sources. The integration of these markets supports the adoption of distributed energy resources (DERs), strengthens grid resilience, and empowers a more active prosumer engagement \cite{gonzalez2022opportunities}. 

Recent studies \cite{strategic1, strategic2, heuristic1, heuristic2, arbitrage1, rule-based2, strategic_marl_2, strategic_marl_3, strategic_marl_4} on strategic bidding in LEMs can be categorized into three main strands: optimization‐based, heuristic/rule‐based, and learning‐based approaches. Bi-level optimization methods formulate a participant’s bidding problem as a leader–follower game, often reformulated as a mathematical program with equilibrium constraints (MPEC) by using equivalent Karush-Kuhn-Tucker (KKT) conditions. For example, \cite{strategic2} developed a bi-level model in which the upper level represents one strategic bidder and the lower level the market‐clearing auction. By relaxing binary variables and applying KKT conditions, the bi-level model was transformed into an MPEC and solved as a single-level program. While this method can deliver exact equilibrium bids, it requires full knowledge of the lower-level market model and other participants' bids, which is impractical in real-world settings. Additionally, it is limited to a single participant and cannot be directly applied to multi-agent systems with multiple market participants. Heuristic and rule-based methods have also been applied in \cite{heuristic1, heuristic2} to maximize participant profits. Unlike bi‐level optimization, they do not guarantee global optima but can efficiently handle the non‐convex, non‐linear, and stochastic problems typical of electricity markets, especially when traditional approaches are computationally intensive or assume full market information. However, heuristic methods only provide a static bidding solution tied to a specific set of market conditions (e.g., prices, loads, generation) and must be rerun whenever these factors change. Therefore, they cannot deliver a strategy that remains valid for future, unseen scenarios. Rule-based methods \cite{arbitrage1, rule-based2} use predefined logic based on market conditions to enable practical, low-complexity strategic bidding in electricity markets. However, their static nature often fails to capture complex market dynamics, leading to suboptimal strategies that typically serve as baselines for more advanced approaches. Learning-based approaches, such as reinforcement learning (RL) algorithms, have been widely employed to simulate bidding strategies, demonstrating their efficacy in dynamic market environments. Unlike bi‐level optimization approaches that provide a single solution and rely on detailed market models, RL is model‐free and learns bidding strategies that generalize across diverse market conditions without full market knowledge. Compared to heuristic or rule-based methods, RL performs better by adapting to dynamic environments and optimizing long-term strategies through data-driven learning. However, RL's computational demands make heuristics more suitable for simpler, well-defined tasks. The authors in \cite{strategic1} introduced DeepBid, a single-agent reinforcement learning framework utilizing Proximal Policy Optimization (PPO) for real-time renewable energy bidding and battery control, leading to increased total profit and operational efficiency. \cite{strategic_marl_1} developed an intelligent strategic bidding framework employing multi-agent transfer learning based on multi-agent RL (MARL) for competitive electricity markets, enhancing market stability and participant profitability. The authors in \cite{strategic_marl_2}, \cite{strategic_marl_3}, and \cite{strategic_marl_4} applied the multi-agent deep deterministic policy gradient (MADDPG) algorithm in double-sided auction LEMs, focusing on both centralized and peer-to-peer (P2P) trading mechanisms. By leveraging public order book information, the MADDPG optimized trading strategies for prosumers and consumers, demonstrating significant economic benefits and operational efficiency. These studies highlight the potential of MARL in enhancing market participant profitability within LEMs.\looseness=-1

\begin{figure*}[!tb]
\centering
    \hspace{0.0cm}
    \includegraphics[scale=0.54]{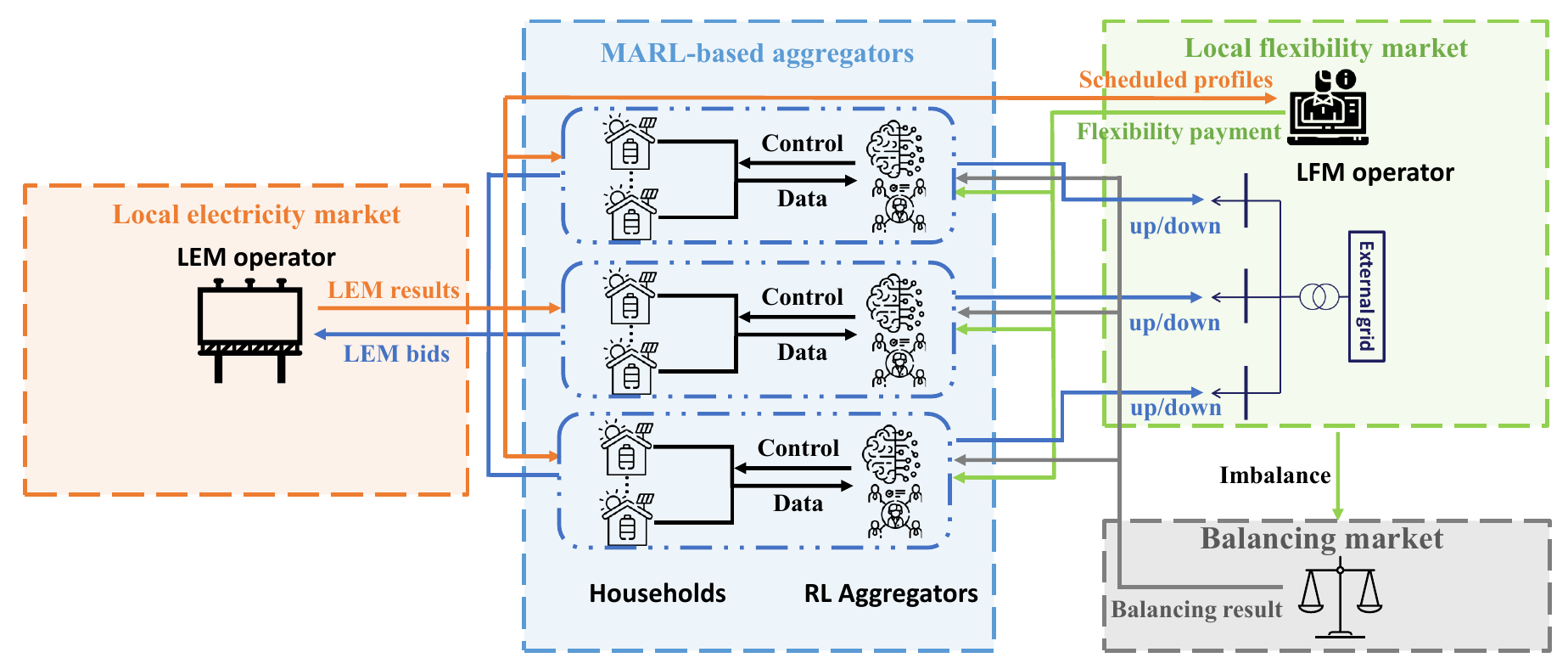}
\caption{Framework of the paper.}
\label{fig:paper_framework}
\end{figure*}

Nevertheless, the aforementioned studies have primarily investigated strategic bidding within a single energy or flexibility market, neglecting arbitrage across multiple markets. While some recent works have begun to explore this area, they often make simplifying assumptions. For example, the work in \cite{arbitrage1} utilized the advantage actor-critic algorithm to simulate arbitrage trading by wholesale market participants across day-ahead, continuous intraday, and balancing markets. Similarly, \cite{arbitrage2} introduced a multi-actor attention-critic method to manage cooperative DERs in a community, aiming to maximize profits in both the LEM and flexibility markets. \cite{arbitrage3} developed a temporal-aware RL approach for battery energy storage systems (BESS) to participate in energy and contingency reserve markets, capitalizing on arbitrage opportunities by exploiting temporal price fluctuations. The temporal-aware RL surpasses traditional RL methods, yielding substantial revenue increases by optimizing bids across multiple markets. 

However, these studies \cite{arbitrage1, arbitrage2, arbitrage3} share two critical shortcomings. First, they have primarily focused on either single or cooperative market participants, neglecting interactions among multiple competitive participants. Second, they often assumed that agents who participate across multiple local markets possess comprehensive information from all relevant markets when making bidding decisions. This assumption simplifies the decision-making process by allowing agents to optimize their strategies based on complete knowledge of market conditions. However, this idealized scenario does not reflect the reality of market operations, where markets are cleared sequentially at multiple stages. For instance, in certain studies, the LFM is typically cleared after the LEM \cite{sequential1, sequential2}. As a result, agents only gain access to the information from the LEM after submitting their bids and once the market has cleared. This sequential and multi-stage market clearing process introduces additional complexity to the decision-making of market participants. Agents must make initial decisions with incomplete information, as they cannot anticipate the outcomes of subsequent market stages until earlier stages are resolved. Therefore, it is crucial to account for the hierarchical and sequential nature of market operations when developing strategies for arbitraging in local markets. This consideration highlights the need for advanced MARL algorithms to effectively model and optimize the decision-making processes in these interconnected and temporally dependent markets. \looseness=-1

The hierarchical MARL (HMARL) algorithm enables the autonomous partitioning of complex multi-stage decision-making tasks into more manageable subtasks, guided by a hierarchy of policies \cite{hmarl1}. The authors in \cite{hmarl2} employed the HMARL algorithm to optimize dispatch control to automatically route and repair multiple repair crews in a coupled power transport network, with the goal of enhancing the resilience of the system against various disturbances. \cite{hmarl3} introduced an HMARL framework to handle high-dimensional action spaces and non-stationary environments. The proposed method demonstrates enhanced coordination among DERs by decomposing control strategies into hierarchical layers, thereby improving the overall system efficiency and reliability in energy dispatch and management tasks. \cite{hmarl4} proposed an HMARL framework for optimizing energy trading strategies within a three-stage Stackelberg game model that encompasses market operators, retailers and consumers in a blockchain-assisted software-defined energy market. This approach enhances decision-making efficiency and economic benefits for all market participants. The research discussed above indicates that the structure of the HMARL algorithm is particularly effective for exploring arbitrage opportunities across various local markets. This hierarchical framework enables effective decision-making and optimization within complex, sequentially-cleared, multi-stage market environments. \looseness=-1

Figure \ref{fig:paper_framework} illustrates the novel HMARL method proposed in this paper, which features two sub-agents enabling aggregators to perform arbitrage across local markets, including the LEM, LFM, and balancing market. Unlike traditional HMARL, where each aggregator employs one critic network and two actor networks, the proposed HMARL assigns each aggregator two sub-agents, a primary sub-agent and a secondary sub-agent, each equipped with its own critic and actor networks. This structure facilitates the simulation of the aggregator's arbitrage and stand-alone (not arbitraging in different markets) trading strategies in local markets, either through siloed or collaborative operations among sub-agents. Firstly, the primary sub-agent of the aggregator strategically bids in the day-ahead LEM to maximize total profit across all local markets. Once the LEM clears, each aggregator’s scheduled dispatch, including consumption or generation profiles, is established for the next day. Subsequently, the secondary sub-agent of the aggregator submits flexibility bids to the LFM to maximize the profit in the LFM. Based on the LEM results and available flexibility bids, the LFM is cleared using an AC optimal power flow model to enforce network constraints by adjusting the day-ahead schedule. Finally, real-time deviations between actual dispatch and the day-ahead schedule are resolved in the balancing market, where aggregators either incur costs or earn revenue based on their imbalance performance. Specifically, the contributions of this paper are as follows: \looseness=-1
\begin{itemize}
\setlength\itemsep{0.0pt}
    \item A framework for local markets is introduced, integrating an LEM for energy trading, an LFM to ensure system operation under grid constraints, and a balancing market to penalize deviations from scheduled profiles. \looseness=-1
    \item The strategic bidding behavior of aggregators in the local markets is formulated as a two-stage Markov game and is solved by the MARL algorithm. This approach offers a model-free method for modeling complex decision-making processes and interactions among the self-interested market participants as a non-cooperative game. \looseness=-1
    \item To capture the bidding behavior of market participants in local markets, a HMARL algorithm, specifically designed for arbitrage, is proposed to solve the two-stage Markov game. Each aggregator employs two sub-agents: one for the first stage and another for the second stage. These sub-agents operate in silos, using two independent MARL algorithms when no arbitrage occurs between markets, and collaborate under the HMARL framework when arbitrage is present.\looseness=-1
\end{itemize}

The structure of the paper is organized as follows: In Section~\ref{sec:LEM}, the structure and formulation of the local markets are presented. The HMARL algorithm is introduced in Section~\ref{sec:hmarl}. Case studies and results analysis are discussed in Section~\ref{sec:result}. Finally, Section~\ref{sec:conclusion} provides the conclusion and suggests directions for future research. \looseness=-1

\section{Local Markets Formulation}\label{sec:LEM}

\begin{figure}[!tb]
\centering
    \hspace{0.0cm}
    \includegraphics[scale=0.28]{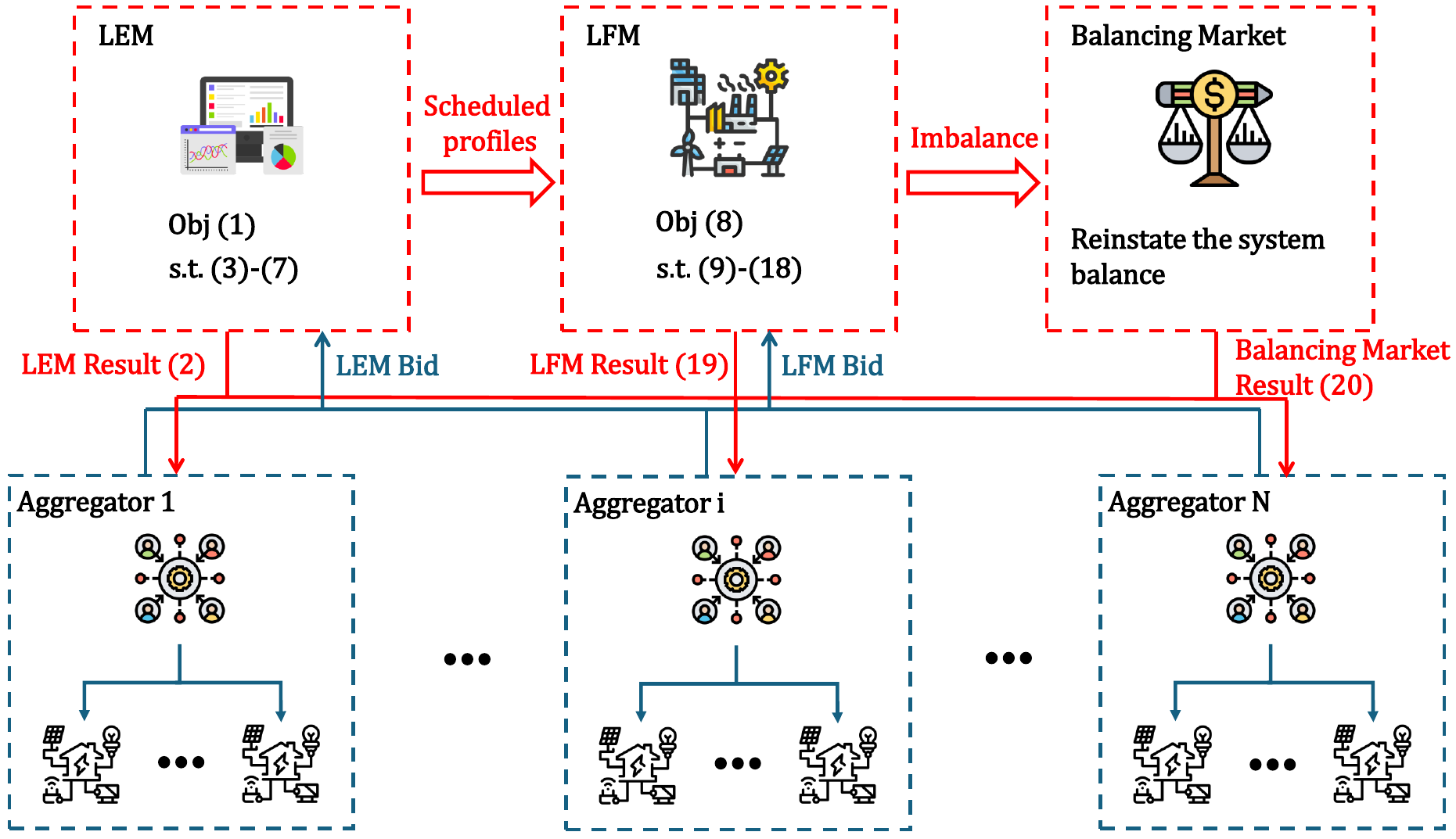}
\caption{Structure of the sequentially-cleared local market.}
\label{fig:market_structure}
\end{figure}

This section introduces the structure and formulation of the sequentially cleared local markets, including the LEM, the LFM, and the balancing market. Market participants are aggregators in the DN that aggregate groups of households and begin by participating in the LEM. After the LEM is cleared, the LFM operator proceeds to clear the market using scheduled profiles and flexibility service bids submitted by aggregators. Subsequently, aggregators also serve as balance responsible parties, bearing the costs incurred by deviations from the scheduled profiles set in the LEM. Figure \ref{fig:market_structure} details the structure of the local markets. \looseness=-1

\subsection{Local Electricity Market}\label{sec:LEM}

The LEM operates as a day-ahead, community-based market \cite{hierarchical2}. Each aggregator $i \in \mathcal{I}$ (where $\mathcal{I}$ is the set of all aggregators) submits the corresponding input variables and parameters for the next day. The input parameters of the LEM are $\{p_{i}^P, p_{t}^{im}, p_{t}^{ex}, g_{i t}^D, \bar{S}^P_{i t}, \bar{g}_{i t}^P, \tan{\theta_{i}}\}$. Here, $p_{i}^P$ signifies the marginal generation cost from energy resources. $p_{t}^{im}$ and $p_{t}^{ex}$ denote the hourly import and export prices, respectively. $g_{i t}^D$ represents the demand. $\bar{S}^P_{i t}$ and $\bar{g}_{i t}^P$ represent the upper bounds for apparent and active power, respectively, of the energy resources. $q_{i t}^P$ denotes the reactive power generation of aggregator $i$ at time $t$. The inclusion of $\tan{\theta_{n}}$ ensures that the energy resources operate at a sufficiently high ratio of reactive to active power, thereby limiting reactive power circulation within the network \cite{pvoperation}. After the market is cleared, the LEM determines market outcomes, and the output control variables are $\{c^{LEM}_i, g_{i t}^P, q_{i t}^P, g_{i t}^{im}, g_{i t}^{ex}\}$, including the aggregators' market cost $c^{LEM}_i$, active power $g^P_{i t}$, and reactive power $q^P_{i t}$ of the generators, and the import/export quantities from the external grid $g_{i t}^{im}/g_{i t}^{ex}$. The market objective is to minimize the total system cost $c^{LEM}$ as follows: \looseness=-1
\begin{equation}
   \begin{split}
        c^{LEM} = {\text{min}}  \sum_{i} c^{LEM}_i\\
    \end{split}
    \label{eq:LEM_1}
\end{equation}
\begin{equation}
   \begin{split}
        c^{LEM}_i =   
        \sum _{t} (p_{i}^P g_{i t}^P
        + p_{t}^{im} g_{i t}^{im}
        - p_{t}^{ex} g_{i t}^{ex})\\
    \end{split}
    \label{eq:LEM_2}
\end{equation}

The energy balance constraint for each aggregator is formulated as follows:\looseness=-1
\begin{equation}
    \begin{split}
        g_{i t}^{im}  + g_{i t}^P =  g_{i t}^{ex} + g_{i t}^D \;,  \forall i \in \mathcal{I}, \forall t \in \mathcal{T}\\
    \end{split}
    \label{eq:LEM_3}
\end{equation}
\begin{equation}
    \begin{split}
        0 \leq g_{i t}^{im},  g_{i t}^{ex} \;,  \forall i \in \mathcal{I}, \forall t \in \mathcal{T}\\
    \end{split}
    \label{eq:LEM_4}
\end{equation}

The operation constraints of the energy sources are as follows:\looseness=-1
\begin{equation}
        (g_{i t}^P)^2 + (q_{i t}^P)^2 \leq (\bar{S}^P_{i t})^2 \;,  \forall i \in \mathcal{I}, \forall t \in \mathcal{T}
    \label{eq:LEM_5}
\end{equation}
\begin{equation}
    \begin{split}
        -\tan{\theta_{i}} g_{i t}^P \leq q_{i t}^P \leq \tan{\theta_{i}} g_{i t}^P \;,  \forall i \in \mathcal{I}, \forall t \in \mathcal{T}
    \end{split}
    \label{eq:LEM_6}
\end{equation}
\begin{equation}
        0 \leq g_{i t}^P \leq \bar{g}_{i t}^P \;,  \forall i \in \mathcal{I}, \forall t \in \mathcal{T}
    \label{eq:LEM_7}
\end{equation}

\subsection{Local Flexibility Market}\label{sec:LFM}

After the LEM clears, aggregators obtain the market results and scheduled power profiles $\hat{g}$. These aggregators act as flexibility service providers and calculate available flexibility quantities based on these profiles. In this paper, the DSO is assumed to be the sole buyer of flexibility services and is responsible for their payment. The grid constraints are formulated as an AC OPF problem based on the linearized distribution flow (LinDistFlow) model \cite{LinDistFlow}. Consider a distribution grid denoted by $\mathcal{DG}=(\mathcal{N},\mathcal{L})$, where $\mathcal{N}$ denotes the set of nodes and $\mathcal{L}$ denotes the set of edges. The aggregators submit flexibility bids to the LFM, and the input parameter of the LFM are $\{p_{n t}^{up}, p_{n t}^{dw}, \hat{g}_{n t}, q_{n t}^D, \bar{S}_n, \bar{S}^{L}_{l}, \tan{\theta_{n}}, \bar{g}_{n}, R_{i n}, X_{i n}, \underline{v}_{n t}, \overline{v}_{n t}\}$. Here, $p_{n t}^{up}$ and $p_{n t}^{dw}$ represent the respective prices associated with the up-regulation and down-regulation bids. $q _{n t}^D$ is the reactive power of the demand. $\bar{S}_n$ and $\bar{S}^L_n$ represent the apparent power limits for nodes and lines, respectively. $\bar{g}_{n}$ denotes the maximum active power limits for each node. $R_{i n}$ and $X_{i n}$ are the resistance and reactance of the line between nodes $i$ and $n$. The voltage magnitudes of the nodes are bounded by $\underline{v}_{n t}$ and $\overline{v}_{n t}$. The LFM operator then clears the market to ensure the aggregators’ scheduled profiles can be physically delivered without violating network constraints, producing the LFM market outcomes as the output, including control variables $\{g_{n t}^{up}, g_{n t}^{dw}, q_{n t}^P\}$ and state variables $\{g_{n t}, q_{n t}, g_{l t}^{L}, q_{l t}^{L}, v_{n t}\}$. $g_{n t}^{up}$ and $g_{n t}^{dw}$ denote the up- and down-regulation quantities of bids submitted by the aggregators at bus $n$ at time $t$. $g_{n t}$ and $q_{n t}$ are the active and reactive power injections at bus $n$ at time $t$, while $g_{l t}^{L}$ and $q_{l t}^{L}$ represent the active and reactive power flow on branch $l$ at time $t$. $v_{n t}$ is the squared voltage magnitues at bus $n$ at time $t$. The objective of the LFM is to minimize the overall cost of flexibility services as follows:\looseness=-1
\begin{equation}
   \begin{split}
        {\text{min}}  \sum _{t} \sum _{n} (p_{n t}^{up} g_{n t}^{up} + p_{n t}^{dw} g_{n t}^{dw} )
    \end{split}
    \label{eq:LFM_1} 
\end{equation}

Given the scheduled profiles from the LEM at each node $\hat{g}_{n t}$, the node power balance constraints are formulated as follows:\looseness=-1
\begin{equation}
    \begin{split}
    	g_{n t} = \hat{g}_{n t} + g_{n t}^{up} - g_{n t}^{dw} \;, \forall n \in \mathcal{N}, \forall t \in \mathcal{T}  \\
    \end{split}
    \label{eq:LFM_2}
\end{equation}
\begin{equation}
    \begin{split}
    	q_{n t} = q_{n t}^P - q_{n t}^D \;, \forall n \in \mathcal{N}, \forall t \in \mathcal{T}  \\
    \end{split}
    \label{eq:LFM_3}
\end{equation}
\begin{equation}
    \begin{split}
    	g_{n t} + \sum _{l \in \delta^{-}(n)} g_{l t}^{L} = \sum _{l \in \delta(n)} g_{l t}^{L} \;, \forall n \in \mathcal{N}, \forall t \in \mathcal{T}  \\
    \end{split}
    \label{eq:LFM_4}
\end{equation}
\begin{equation}
    \begin{split}
    	q_{n t} + \sum _{l \in \delta^{-}(n)} q_{l t}^{L} = \sum _{l \in \delta(n)} q_{l t}^{L} \;, \forall n \in \mathcal{N}, \forall t \in \mathcal{T}  \\
    \end{split}
    \label{eq:LFM_5}
\end{equation}

Here, $\delta^{-}(n)$ denotes the set of lines entering node $n$, and $\delta^{+}(n)$ denotes the set of lines exiting node $n$. The power limits of the nodes and branches are written as:\looseness=-1
\begin{equation}
        (g_{n t})^2 + (q_{n t})^2 \leq (\bar{S}_n)^2, \forall n \in \mathcal{N}, \forall t \in \mathcal{T} \\
    \label{eq:LFM_6}
\end{equation}
\begin{equation}
        (g_{l t}^{L})^2 + (q_{l t}^{L})^2 \leq (\bar{S}^{L}_{l})^2 , \forall  l \in \mathcal{L}, \forall t \in \mathcal{T}
    \label{eq:LFM_7}
\end{equation}
\begin{equation}
    \begin{split}
        -\tan{\theta_{n}} g_{n t} \leq q_{n t} \leq \tan{\theta_{n}} g_{n t} \;,  \forall n \in \mathcal{N}, \forall t \in \mathcal{T}
    \end{split}
    \label{eq:LEM_8}
\end{equation}
\begin{equation}
    \begin{split}
    	| g_{n t} | \leq \bar{g}_{n} \;, n \in \mathcal{N}, t \in \mathcal{T}  \\
    \end{split}
    \label{eq:LFM_9}
\end{equation}

The relationship between the voltage magnitudes squared, denoted by $v_{n t}$ and $v_{i t}$ of connected nodes $n$ and $i$, is given by:\looseness=-1
\begin{equation}
    \begin{split}
        v_{n t} = v_{i t} - 2 R_{i n} g_{i n t}^{L} &- 2 X_{i n} q_{i n t}^{L} \\
     \;, & n \in \mathcal{N}, i \in \delta^{-}(n), t \in \mathcal{T} \\
    \end{split}
    \label{eq:LFM_10}
\end{equation}

And the voltages are bounded as:\looseness=-1
\begin{equation}
    \begin{split}
    	\underline{v}_{n t} \leq v_{n t} \leq \overline{v}_{n t} \;, n \in \mathcal{N}, t \in \mathcal{T} \\
    \end{split}
    \label{eq:LFM_11}
\end{equation}

Similar to the locational marginal price (LMP) used in the national wholesale electricity market, the distributed LMP (DLMP) provides price signals at the nodes of the distribution grid. Following works in \cite{DLMP1, DLMP2}, the dual variable $\eta^{g}_{n t}$, representing the Lagrange multiplier associated with the active power balance constraint defined in (\ref{eq:LFM_4}), serves as the DLMP. This indicates the marginal cost of flexibility due to a one-unit increase in scheduled active power injection at the node. The net cost $c^{FLX}_i$ at node $n$ incurred by aggregators for providing flexibility services is calculated as:\looseness=-1
\begin{equation}
    \begin{split}
        c^{FLX}_i = (\eta^{g}_{n t} - p_{n t}^{up}) p_{n t}^{up} + (\eta^{g}_{n t} - p_{n t}^{dw}) p_{n t}^{dw} \\
    \end{split}
    \label{eq:LFM_12}
\end{equation}

\subsection{Balancing Market}\label{sec:BAL}

After an LEM clears, aggregators receive their scheduled profiles. Subsequent deviations occur after the LFM clearing, where aggregators provide up-regulation or down-regulation flexibility services, necessitating financial settlements from the balancing market. If the deviation is positive, meaning the aggregator's net generation exceeds the scheduled profiles, the aggregator is compensated by the balancing market at the positive balancing price $p_{t}^{BAL,pos}$. Conversely, if the deviation is negative, indicating the net generation falls short of the scheduled profiles, the aggregator must pay the balancing market at the negative balancing price $p_{t}^{BAL,neg}$. The final financial settlement for aggregators with the balancing market, denoted by $c^{BAL}_i$, is calculated as:\looseness=-1
\begin{equation}
    \begin{split}
    c^{BAL}_{i t} &= \begin{cases}
     p_{t}^{BAL,neg} (\hat{g}_{i t} - g_{i t}) & \text{if } g_{i t} \leq \hat{g}_{i t}, \\
    p_{t}^{BAL,pos} (g_{i t} - \hat{g}_{i t}) &  \text{otherwise},
    \end{cases}
    \label{eq:BAL_1}
    \end{split}
\end{equation}

\section{Bidding Strategy with Hierarchcial Multi-agent Reinforcement learning}\label{sec:hmarl}

In this section, the strategic bidding behavior of market participants in the local markets is formulated as a two-stage Markov game with partially observable agents, with the market clearing process shown in Figure \ref{fig:market_structure}. The first stage captures decision-making in the LEM, while the second stage addresses decisions in the LFM and balancing market. An HMARL algorithm is then designed to solve the problem, allowing aggregators to arbitrage across the local markets. \looseness=-1

\subsection{Problem Formulation as a Two-stage Markov Game}\label{sec:POMG}

\begin{figure*}[!tb]
\centering
    \hspace{0.0cm}
    \includegraphics[scale=0.43]{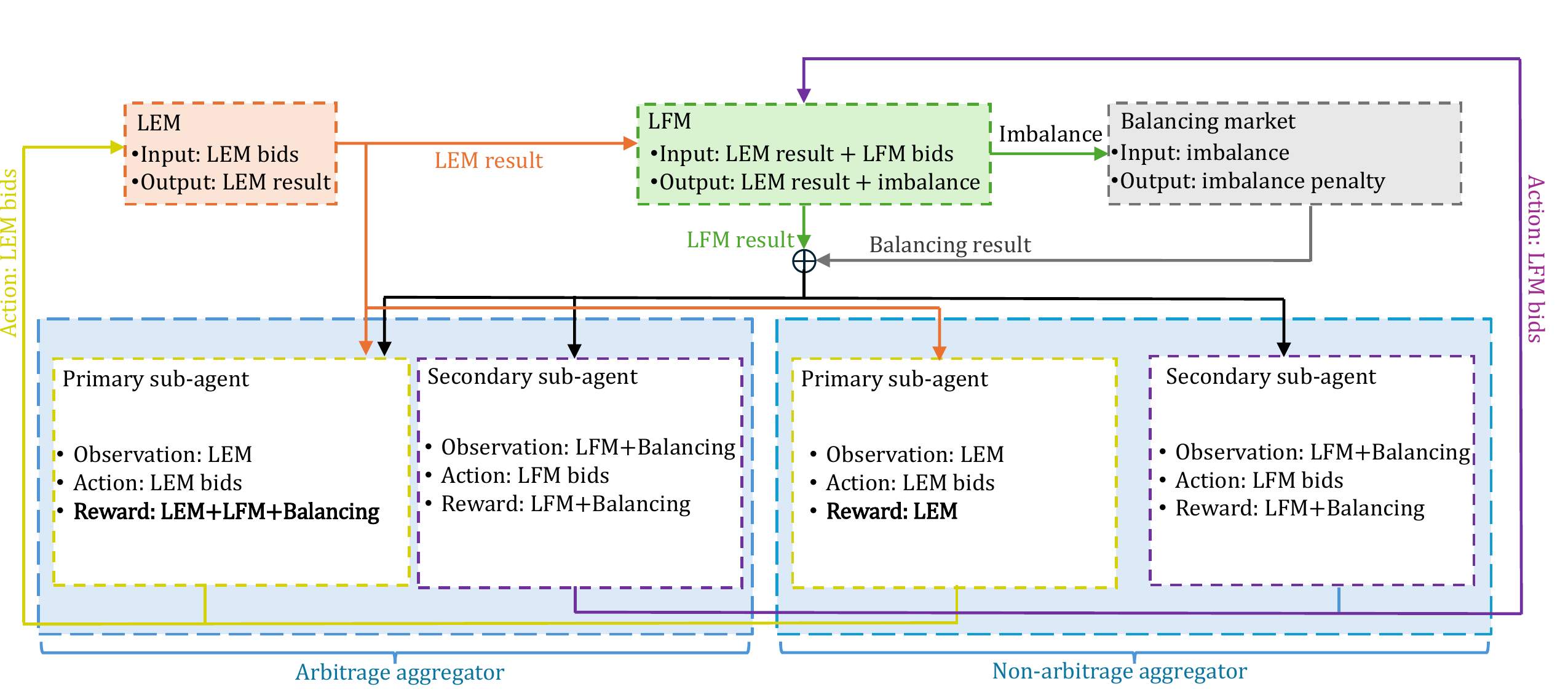}
\caption{Two-stage Markov game.}
\label{fig:markov_game}
\end{figure*}

The strategic bidding behavior of self-interested aggregators in a single LEM or LFM is typically formulated as a Markov game \cite{MDP, strategic_marl_1, strategic_marl_2}. In this framework, aggregators make strategic bids, $\boldsymbol{a}_{t} = (a_{1 t}, a_{2 t},..., a_{N t})$, based on their observations, $\boldsymbol{o}_{t} = (o_{1 t}, o_{2 t},..., o_{N t})$ as part of the current state of the market $s_t$, at time step $t$. The bidding policies, $\boldsymbol{\pi} = (\pi_1, \pi_2,...,\pi_N)$, determine the bids such that $a_{i t} = \pi_i(o_{i t})$ for all $ i \in \mathcal{I}$. After the market clears, the aggregators receive net cost reduction represented by rewards $\boldsymbol{r}_{t} = (r_{1 t}, r_{2 t},..., r_{N t})$. The market then transits to the next state, $s_{t+1}$ with the a probability represented as $p\bigl(s_{t+1}|s_t,\boldsymbol{a}_{t}\bigl)$. Aggregators aim to develop trading policies to maximize their discounted cumulative reward $G_{i t} = \sum_{t}^{T} \gamma^{t-1} r_{i t}$ by employing a discount factor $\gamma \in [0,1]$ that balances the importance of immediate rewards versus future rewards. The discount factor determines the relative weight placed on future rewards, with a value closer to 1 signifying a greater emphasis on long-term benefits. As the aggregator makes decisions, it estimates the expected future profits using the state-value function $V_i(\boldsymbol{o}_t) = \mathbf{E}_{\pi_i}(G_{it}|\boldsymbol{o}_t)$. 

However, in our proposed local markets framework, aggregators make sequential decisions across multiple markets at each time step based on the observation from each market, distinguishing it from traditional Markov games, which involve only a single observation, action, and reward at each time step. During the bidding process, aggregators solely rely on observations from the LEM to submit bids. Only after the LEM clears do aggregators receive observations of the LFM and the balancing market, allowing them to submit bids to the LFM. To address this, a two-stage Markov game is introduced, where each aggregator takes action in two steps. In the first stage, the aggregator submits its bid to the LEM. After the LEM clears, the aggregator, informed by the LEM outcome, submits a bid in the LFM considering both the anticipated income from the LFM and the imbalance penalties incurred in the balancing market. In this case, the observation in the LEM of the first-stage, denoted as $o^{LEM}_{i t}=[t, p^{net}_{i t}, p^{im}_t]$, includes the current time step $t$, net generation $p^{net}_{i t}$, and the hourly prices in the LEM $p^{im}_t$. Subsequently, the aggregator takes action $a^{LEM}_{i t}$, a value between 0 and 1, under the policy $\pi^{LEM}_{i}$ that signifies the amount withheld from the bid quantity of energy resources. Consequently, the final bid quantity of energy resources submitted to the LEM is $a^{LEM}_{i t} g^P_{i t}$. After the LEM is cleared, the aggregator receives the reward as follows:\looseness=-1
\begin{equation}
    \begin{split}
        r^{LEM}_{i t} = \bar{c}^{LEM}_{i t} -c^{LEM}_{i t}\\
    \end{split}
    \label{eq:reward1}
\end{equation}
where the reward function $r^{LEM}_{i t}$ for the aggregator at time step $t$ captures the cost reduction achieved through strategic physical withholding, which involves intentionally withholding a portion of its actual generation capacity from the market. This reward is calculated as the difference between the aggregator's actual net cost in the LEM $c^{LEM}_{i t}$ defined in (\ref{eq:LEM_2}) and a constant reference cost $\bar{c}^{LEM}_{i t}$ representing the net cost incurred under truthful bidding without withholding quantity. 

After obtaining the LEM results, the aggregator receives an observation denoted as $o^{LFM}_{i t}=[t, p^{net}_{i t}, a^{LEM}_{i t}, p_{t}^{BAL,pos}, p_{t}^{BAL,neg}]$. Following the policy $\pi^{LFM}_{i}$, it takes the actions that represent the bidding prices for up-regulation and down-regulation services, given by $a^{LFM}_{i t}=[p_{i t}^{up}, p_{i t}^{dw}]$. After the market operator receives all flexibility bids and clears the LFM, the aggregator receives the reward $r^{LFM}_{i t}$ defined as:\looseness=-1
\begin{equation}
    \begin{split}
    	r^{LFM}_{i t} = -c^{FLX}_{i t} -c^{BAL}_{i t}\\
    \end{split}
    \label{eq:reward2}
\end{equation}
where $c^{FLX}_{i t}$ and $c^{BAL}_{i t}$ represent the net costs incurred by the aggregator in the second stage, including the LFM and the balancing market, defined in (\ref{eq:LFM_12}) and (\ref{eq:BAL_1}), respectively. \looseness=-1

\subsection{Hierarchical Multi-agent Deep Deterministic Policy Gradient}\label{sec:maddpg}

To solve the two-stage Markov game, a model-free MARL algorithm with centralized training and decentralized execution, namely hierarchical multi-agent deep deterministic policy gradient (HMADDPG), is utilized to account for the arbitrage behaviors of the aggregators, as shown in Figure \ref{fig:markov_game}. MARL typically derives optimal strategies using the Q-value function, defined as $Q_{i} (\boldsymbol{o}_t, \boldsymbol{a}_t) = \mathbf{E}_{\pi_i}(G_{it}|\boldsymbol{o}_t, \boldsymbol{a}_t)$. The Q-value function can be expressed in a recursive format according to the Bellman equation as follows:\looseness=-1
\begin{equation}
    \begin{split}
        Q_i(\boldsymbol{o}_t, \boldsymbol{a}_t) = \mathbf{E}(r_{i t}+\gamma Q_i(\boldsymbol{o}_{t+1}, \boldsymbol{a}_{t+1})) \\
    \end{split}
    \label{eq:bellman_1}
\end{equation}  

Consequently, the Q-value function can be updated using the bootstrapping method, which allows updating future Q-value estimates based on current estimates. The Temporal Difference (TD) method learns directly from experience without requiring a model of the environment. A key advantage of TD methods over Monte Carlo methods is their ability to update estimates using previously learned Q-values without requiring the final outcome, as shown below:\looseness=-1
\begin{equation}
    \begin{split}
        Q_i (\boldsymbol{o}_t, \boldsymbol{a}_t) = Q_i (\boldsymbol{o}_t, \boldsymbol{a}_t) + \alpha (r_{i t}+ \\
        \gamma \max_{a_{i t+1}} Q_i (\boldsymbol{o}_{t+1}, \boldsymbol{a}_{t+1})&
        -Q_i (\boldsymbol{o}_t, \boldsymbol{a}_t)) \\
    \end{split}
    \label{eq:bellman_2}
\end{equation} 
where $\alpha \in [0,1]$ is the step size \cite{rl1}. \looseness=-1

In terms of aggregators with different trading strategies, if the aggregator $\boldsymbol{i}^{sa} \in \mathcal{I}^{sa}$ (where $\mathcal{I}^{sa}$ is the set of stand-alone aggregator) does not arbitrage between local markets, its two sub-agents $(i^{sa}_{PR}, i^{sa}_{SE} \in \boldsymbol{i}^{sa})$ operate in silos without any communication between them. The primary sub-agent $i^{sa}_{PR}$ aims to maximize its Q-value function $Q^{PR}_{i^{sa}_{PR}}(\boldsymbol{o}^{PR}_t,\boldsymbol{a}^{PR}_t)$ which takes as input a joint observation encompassing information solely from the LEM. This joint observation, denoted as  $\boldsymbol{o}^{PR}_t = \boldsymbol{o}^{LEM}_t$, includes the current time step $t$, the joint net generation of the aggregators $\boldsymbol{p}^{net}$, and the hourly import prices $p^{im}$ within the LEM. And the joint action of the aggregators comprises all LEM bids of all primary sub-agents as $\boldsymbol{a}^{PR}_t = \boldsymbol{a}^{LEM}_t$. This information allows the Q-value function to estimate the expected cumulative rewards from the LEM, with the reward $r_{i^{sa}_{PR}}^{LEM}$. After obtaining the LEM results, the secondary sub-agent $i^{sa}_{SE}$ of the aggregator $\boldsymbol{i}^{sa}$ maximize the Q-value function $Q^{SE}_{i^{sa}_{SE}}(\boldsymbol{o}^{SE}_t,\boldsymbol{a}^{SE}_t)$ takes as input a joint observation $\boldsymbol{o}^{SE}_t = \boldsymbol{o}^{LFM}_t$ encompassing information solely from the LFM. The joint action includes all LFM bids of all secondary sub-agents as $\boldsymbol{a}^{SE}_t = \boldsymbol{a}^{LFM}_t$. This information allows the Q-value function to estimate the expected cumulative rewards from the LFM, with the reward $r_{i^{sa}_{SE}}^{LFM}$. \looseness=-1

For the aggregator $\boldsymbol{i}^{arb} \in \mathcal{I}^{arb}$ (where $\mathcal{I}^{arb}$ is the set of arbitrage aggregator) arbitrages in local markets, its two sub-agents $(i^{arb}_{PR}, i^{arb}_{SE} \in \boldsymbol{i}^{sa})$ operate as two collaborative agents. Unlike stand-alone aggregators, where primary sub-agents focus solely on maximizing LEM rewards, the primary sub-agent $i^{arb}_{PR}$ in an arbitrage aggregator submits bids in the LEM to maximize the total reward across both market stages with the reward $r^{PR}_{i^{sa}_{PR} t}$ defined as follows:
\begin{equation}
    \begin{split}
        r^{PR}_{i^{arb}_{PR} t} = -c^{LEM}_{i^{arb}_{PR} t} -c^{FLX}_{i^{arb}_{SE} t} -c^{BAL}_{i^{arb}_{SE} t} \\
    \end{split}
    \label{eq:reward3}
\end{equation} \looseness=-1

This approach considers that LEM bids affect both the LEM market results in the first-stage Markov game and subsequent decisions in the LFM and the balancing market in the second-stage Markov game, ultimately contributing to the aggregator's overall profitability. During training, the primary sub-agent's samples experience tuples—comprising observations, actions, and rewards for both sub-agents—from a replay buffer of past interactions, eliminating the need for future information. The secondary sub-agent $i^{arb}_{SE}$, however, shares the same objective as those in stand-alone aggregators, aiming to maximize their cumulative rewards from the LFM and balancing market, since the actions in the second-stage Markov game only influence the market result of the LFM and the balancing market. 

Compared to traditional MADDPG, where the aggregator utilizes a single critic network and actor network, the proposed HMADDPG introduces a hierarchical structure for the aggregator, where each sub-agent is equipped with a sub-critic network to estimate the Q-value. The Q-values for sub-agents are represented by critic networks $Q^{PR}_i(\boldsymbol{o}^{PR}_t, \boldsymbol{a}^{PR}_t | \theta_i^{Q,PR})$ for primary sub-agents and $Q^{SE}_i(\boldsymbol{o}^{SE}_t, \boldsymbol{a}^{SE}_t | \theta_i^{Q,SE})$ for secondary sub-agents, where $\theta_i^{Q,PR}$ and $\theta_i^{Q,SE}$ denote the respective critic network parameters. Similarly, each sub-agent has a sub-actor network for generating bids. The primary sub-agent's policy is denoted by $\mu^{PR}_i (o^{PR}_{i t}|\theta_i^{\mu,PR})$ for the LEM, and the secondary sub-agent's policy is denoted by $\mu^{SE}_i (o^{SE}_{i t}|\theta_i^{\mu,SE})$ for the LFM and the balancing market, where $\theta_i^{\mu,PR}$ and $\theta_i^{\mu,SE}$ denote the respective actor network parameters. To ensure stable learning and consistent target values during TD updates, supplementary target networks are employed alongside the primary sub-agent critic and actor networks. The primary and secondary sub-agents' target critic networks, denoted by $\hat{Q}^{PR}_i(\boldsymbol{o}^{PR}_t, \boldsymbol{a}^{PR}_t|\theta_i^{\hat{Q},PR})$ and $\hat{Q}^{SE}_i(\boldsymbol{o}^{SE}_t, \boldsymbol{a}^{SE}_t|\theta_i^{\hat{Q},SE})$ respectively, and their corresponding target actor networks, represented by
$\hat{\mu}^{PR}_i (o^{PR}_{i t}|\theta_i^{\hat{\mu},PR})$ and $\hat{\mu}^{SE}_i (o^{SE}_{i t}|\theta_i^{\hat{\mu},SE})$, share the same architecture as their primary counterparts. Their parameters are periodically updated through soft copying from the primary networks. The target networks are updated as follows:\looseness=-1
\begin{equation}
    \begin{split}
        \theta_i^{\hat{Q},b} = \tau \theta^{Q,b}_i + (1-\tau)\theta_i^{\hat{Q},b}, \forall b \in \{PR, SE\}  \\
    \end{split}
    \label{eq:ddpg1}
\end{equation} 
\begin{equation}
    \begin{split}
        \theta_i^{\hat{\mu},b} = \tau \theta_i^{\mu, b} + (1-\tau)\theta_i^{\hat{\mu},b}, \forall b \in \{PR, SE\}  \\
    \end{split}
    \label{eq:ddpg2}
\end{equation}
where $\tau$ is the learning rate close to 1. To train the networks, two separate reply buffers, $\mathcal{R}_i^{PR}$ and $\mathcal{R}_i^{SE}$, are created for each sub-agent $b \in \{PR, SE\}$ of the aggregator $i$. These buffers store historical data from the local markets. During training, a mini-batch of size $N^{m}$ is uniformly sampled from each reply buffer. The parameters of the critic and actor networks for each sub-agent are then updated. The sub-agents of the aggregator minimizes the Smooth L1 Loss function $L(\theta^{Q,b}_i)$, to update its critic network parameters $\theta^{Q,b}_i$ as follows:\looseness=-1
\begin{align}
	L(\theta^{Q,b}_i) &= \begin{cases}
	\frac{1}{N^{m}} \sum_{n=1}^{N^{m}} 0.5 \zeta_{n}^{2} & \text{if } |\zeta_{n}| \leq 1,  \\
	\frac{1}{N^{m}} \sum_{n=1}^{N^{m}} (|\zeta_{n}| - 0.5) &  \text{otherwise},
	\end{cases}
	\label{eq:ddpg3}
\end{align}
\begin{equation}
    \begin{split}
        \zeta_{n} = r^{b}_{n t} + \gamma \hat{Q}^{b}_n &(\boldsymbol{o}^{b}_{t+1}, \boldsymbol{a}^{b}_{t+1}|\theta^{\hat{Q},b}_n)\\
        &-Q^{b}_n (\boldsymbol{o}^{b}_t, \boldsymbol{a}^{b}_t|\theta^{Q,b}_n), \forall b \in \{PR, SE\}  \ \\
    \end{split}
    \label{eq:ddpg4}
\end{equation} 

The Smooth L1 Loss function aims to minimize the TD error, denoted as $\zeta_n$, within the mini-batch. This error represents the difference between the estimated Q-values of the target and regular networks. The neural network parameters are then updated through backpropagation. The actor network is designed to optimize bids by maximizing the Q-value formalized as $J(\theta^{\mu, b}_i)$ in (\ref{eq:ddpg5}). The network parameters are adjusted based on the policy gradient theorem outlined in (\ref{eq:ddpg6}), as described below:\looseness=-1
\begin{equation}
    \begin{split}
        J(\theta^{\mu,b}_i) = \mathbf{E}(Q_i (\boldsymbol{o}^b_t, \boldsymbol{a}^b_t(\theta^{\mu,b}_i)|\theta^{Q,b}_i), \forall b \in \{PR, SE\}   \\
    \end{split}
    \label{eq:ddpg5}
\end{equation} 
\begin{equation}
    \begin{split}
        \nabla_{\theta^{\mu,b}_i} J(\theta^{\mu,b}_i) = \frac{1}{N^{m}} \sum_{n=1}^{N^{m}} (\nabla_{a^b_{i t}}  Q^b_i(\boldsymbol{o}^b_t, \boldsymbol{a}^b_t(\theta^{\mu,b}_i)| \theta^{Q,b}_i)  \\ 
        \nabla_{ \theta^{\mu,b}_i} a^b_{i t}(\theta^{\mu,b}_i)), \forall b \in \{PR, SE\}    \\
    \end{split}
    \label{eq:ddpg6}
\end{equation} 

To encourage exploration and prevent convergence to local optima, Gaussian noise, denoted as $\N_{i t}^{\epsilon}$, is added to the actions of the sub-agents as follows:\looseness=-1
\begin{equation}
    \begin{split}
        a^b_{i t} = clip( \mu_i(o^b_{i t}|\theta^{\mu,b}_i) + \N^{\epsilon}_{i t}(0, \sigma^{\epsilon}_i), a^{b,min}_{i t}, a^{b,max}_{i t}) \\
    \end{split}
    \label{eq:ddpg7}
\end{equation} 
where $\sigma^{\epsilon}_i$ represents the standard deviation of the Gaussian noise. To mitigate excessive exploration and promote convergence over time, $\sigma^{\epsilon}_i$ is subject to linear decay across episodes. This decay is achieved by multiplying the parameter by a decay factor $\kappa$ at the end of each episode. The primary distinctions between the arbitrage and stand-alone aggregators lie in the definition of rewards and input observations, as elaborated in Section \ref{sec:POMG}. Under the stand-alone strategy, the aggregators utilize two independent MADDPG algorithms to make the bidding decisions in the local markets. The details of the HMADDPG algorithm for both types of aggregators are detailed in Algorithm \ref{alg:marl}. \looseness=-1 
\begin{algorithm}[t]
\caption{HMADDPG Algorithms}\label{alg:marl}
\begin{algorithmic}[1]
\State  Create reply buffer $\mathcal{R}$ and configure hyperparameters
\State Initialize regular (target) actor and critic networks for the sub-agents of the aggregators $\theta^{\mu,PR}_i$ ($\theta^{\hat{\mu}, LD}_n$), $\theta^{\mu,SE}_i$ ($\theta^{\hat{\mu},SE}_i$), $\theta_i^{Q,PR}$ ($\theta_i^{\hat{Q},PR}$), $\theta_i^{Q,SE}$ ($\theta_i^{\hat{Q},SE}$)  
\While{$episode \leq N^E$}
    \State Primary sub-agents of the aggregators acquire initial \Statex \hspace{1.1em} observations of the LEM $\boldsymbol{o}^{PR}_{i t}$ at $t = 1$
    \While{$t \leq T$}
        \State Primary sub-agents submit LEM bids $\boldsymbol{a}^{PR}_{t}$
        \State Clear LEM (\ref{eq:LEM_1})-(\ref{eq:LFM_9})
        \State Stand-alone aggregators' primary sub-agents \Statex \hspace{2.7em} obtain rewards $\boldsymbol{r}^{PR}_{t}$ (\ref{eq:reward1}) 
        \State Secondary sub-agents acquire the observation $\boldsymbol{o}^{SE}_{t}$
        \State Secondary sub-agents submit LFM bids $\boldsymbol{a}^{SE}_{t}$
        \State Clear LFM and balancing market (\ref{eq:LFM_1})-(\ref{eq:LFM_11})
        \State Aggregators' secondary sub-agents \Statex \hspace{2.7em} obtain rewards $\boldsymbol{r}^{SE}_{t}$ (\ref{eq:reward2})
        \State Arbitrage aggregators' primary sub-agents \Statex \hspace{2.7em} obtain rewards $\boldsymbol{r}^{PR}_{t}$ (\ref{eq:reward3})   
        \State Add historical data into the reply buffers
        \State Update critic networks (\ref{eq:ddpg3}) and (\ref{eq:ddpg4})
        \State Update actor network by (\ref{eq:ddpg5}) and (\ref{eq:ddpg6})
        \State Soft update target networks by (\ref{eq:ddpg1}) and (\ref{eq:ddpg2})
        \State $t \leftarrow t + 1$
    \EndWhile
\State $\sigma^{\epsilon}_n \leftarrow \kappa \sigma^{\epsilon}_n$ for each aggregator $i$
\State $episode \leftarrow episode + 1$
\EndWhile\label{euclidendwhile}
\State End training
\end{algorithmic}
\end{algorithm}

\section{Case Study}\label{sec:result}

\subsection{Experimental Setup}
\begin{figure}[!tbh]
	\centering
		\includegraphics[width=0.56\linewidth,trim=2 2 2 2,clip]{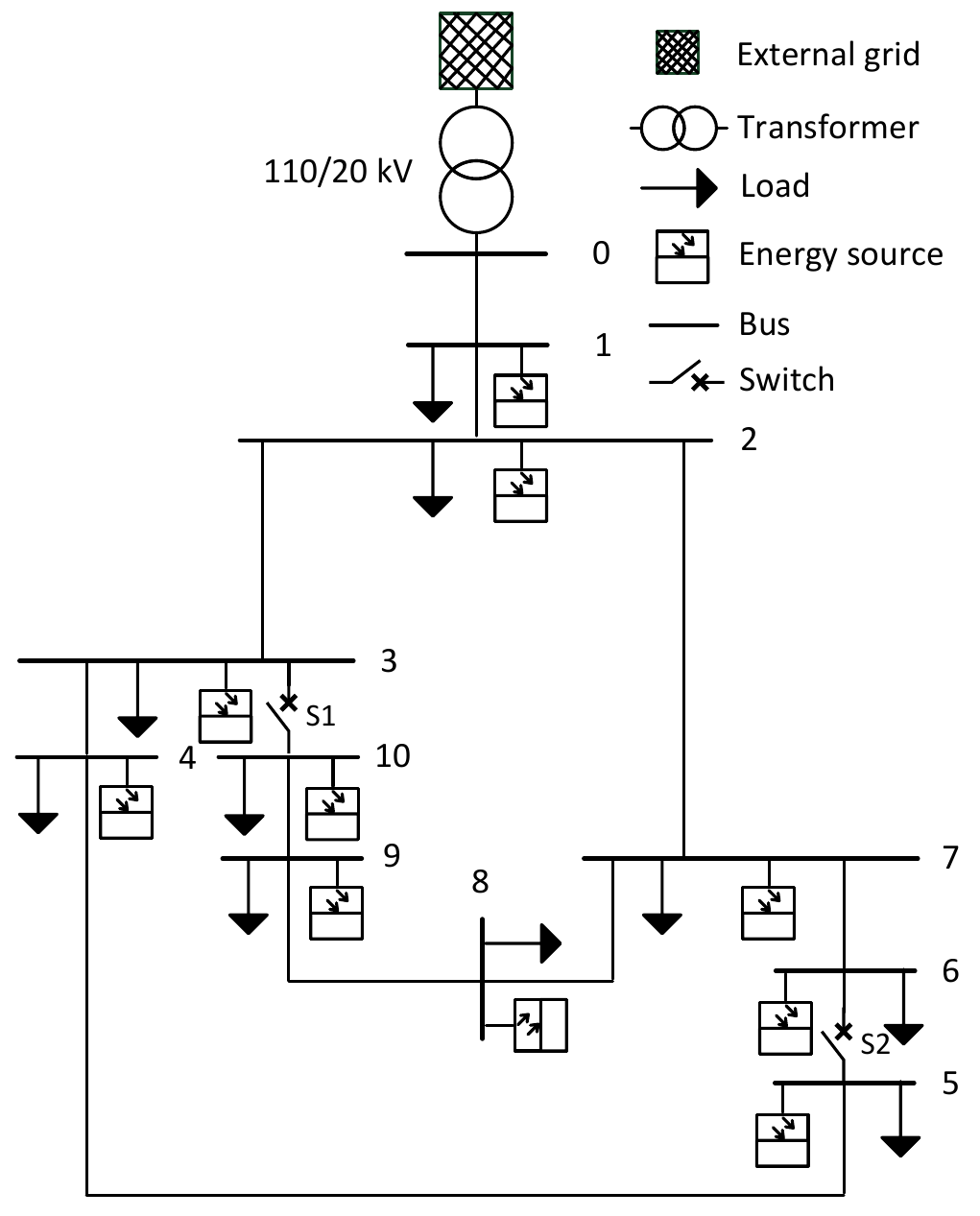}
	\caption{Test system.}
	\label{fig:test_grid}
\end{figure}

This section evaluates the performance of the HMARL algorithm within the proposed sequentially cleared local markets. Figure \ref{fig:test_grid} depicts the 20 kV, 11-bus MV-DN network under investigation. Bus $1$ serves as the system's slack bus, and the base power is set at 1 MVA. Assuming a balanced system, a single-phase model is employed. To facilitate radial operation, switches at branches $(3, 10)$ and $(5, 6)$ are maintained in the open position. Detailed grid parameters can be found in \cite{testgrid}. The DN comprises of ten aggregators. Each aggregator manages the energy resources and demands of a specific bus, excluding the slack bus, and is responsible for coordinating energy trading, flexibility services, and imbalancing. \looseness=-1

The dataset covers an one-year period, consisting of 52 weeks, with 39 weeks allocated for training, 1 week for evaluation, and 12 weeks reserved for testing. The test weeks are evenly sampled across the year with one test set for every four consecutive weeks. For each episode, a random training week is selected from the 39 available. Following MARL training, the aggregators are evaluated on the evaluation week to generate the training curve. Upon reaching the episode limit, the trained aggregators are tested on the 12 designated test weeks. Hourly electricity prices in the LEM, along with positive and negative imbalance prices in the balancing market, are collected from $1^{st}$ January 2020 to $1^{st}$ January 2021, in the Netherlands using data obtained from ENTSO-E \cite{hourlyprice}. A price spread factor of 0.9 is adopted for hourly import and export LEM prices, in accordance with the works presented in \cite{pricefactor}. The hyperparameters of the MARL algorithm are determined using the grid search method. Critic networks of the sub-agents employ a discount factor of 0.99. The minibatch size is set to 35, while the reply buffer accommodates $5\times10^{4}$ samples. Both critic and actor networks of the sub-agents are optimized using the Adam optimizer with learning rates of $5\times10^{-4}$ and $10^{-4}$, respectively. Target networks are softly updated with a rate of 0.001 undergoing five updates per episode. Exploration is facilitated through Gaussian noise with a mean of 0 and a standard deviation of 0.3. The MARL algorithms are implemented in Pytorch 1.12.1 \cite{pytorch}, and the optimization model is solved using the Gurobi 9.5.2 solver \cite{gurobi} on a 6-core 2.60 GHz Intel(R) Core(TM) i7-9750H CPU system equipped with 16 GB of RAM.  \looseness=-1

\subsection{MARL Result Analysis}

\begin{figure*}[bt]
\centering
    \hspace{-0.0cm}
    \includegraphics[scale=1.10]{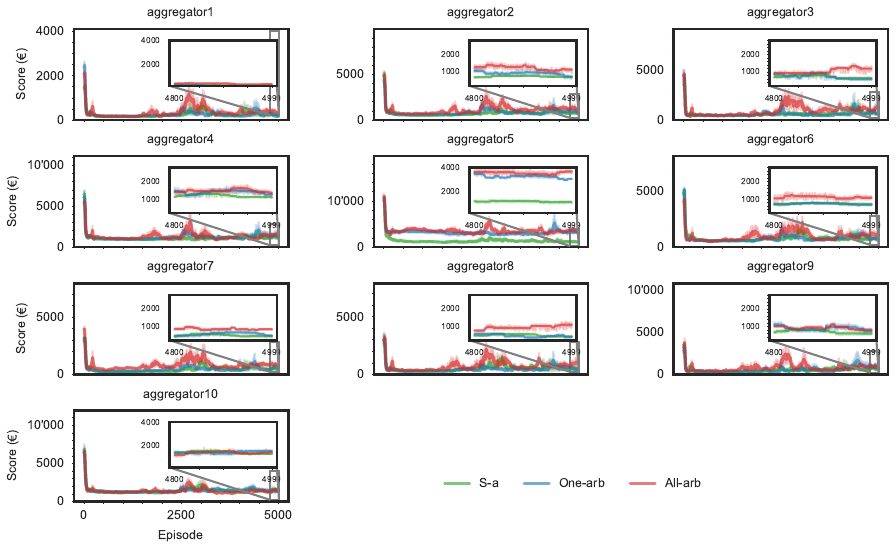}
\caption{Convergence of the aggregators on the evaluation week.}
\label{fig:converge_agent}
\end{figure*}

\begin{figure}[bt]
\centering
    \hspace{-0.0cm}
    \includegraphics[scale=0.49]{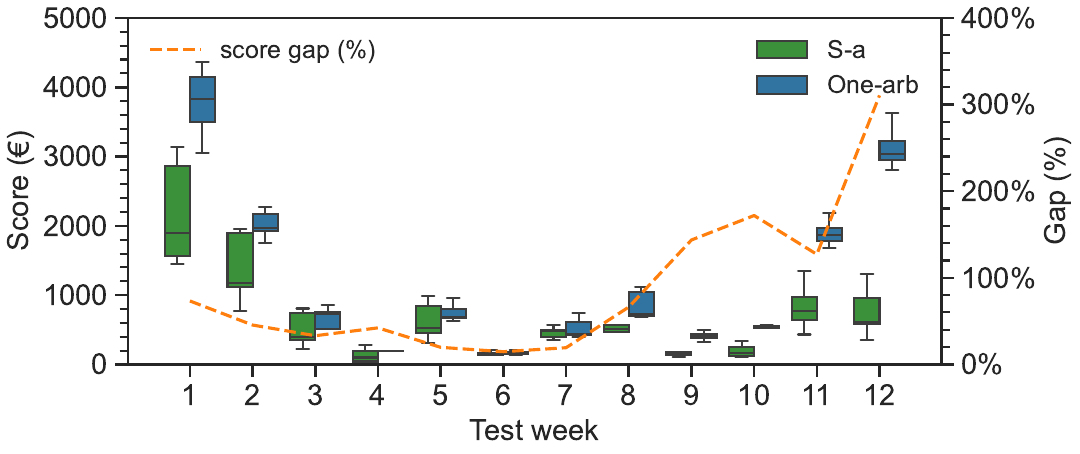}
\caption{Scores of the aggregator 5 during the test weeks.}
\label{fig:score_box}
\end{figure}

\begin{table*}[h!bt]
\caption{Scores of the arbitrage aggregator 5 during the test weeks}
\begin{center}
    \label{tab:profit}
    \renewcommand\arraystretch{1.0}
    \setlength{\tabcolsep}{1.0mm}{
\begin{tabular}{llllllllllllll}
\hline
\multicolumn{2}{l}{Test week}                                           & 1      & 2      & 3     & 4     & 5     & 6     & 7     & 8     & 9      & 10    & 11     & 12     \\ \hline
\multicolumn{1}{c}{\multirow{3}{*}{Stand-alone}} & First-stage score (€) & -8.5   & -0.5   & 0.0   & 0.0   & 0.0   & -0.2  & -1.7  & -0.8  & -4.1   & -1.3  & -2.4   & -37.6  \\
\multicolumn{1}{c}{}                               & Second-stage score (€) & 2189.9 & 1387.6 & 502.5 & 127.6 & 625.4 & 147.5 & 442.3 & 516.1 & 173.4  & 198.8 & 837.1  & 799.3  \\
\multicolumn{1}{c}{}                               & Total score (€) & 2181.4 & 1387.0 & 502.4 & 127.6 & 625.4 & 147.4 & 440.6 & 515.3 & 169.3  & 197.5 & 834.7  & 761.7  \\ \hline
\multirow{5}{*}{One-arbitrage}                         & First-stage score (€) & -181.6 & -15.5  & -2.0  & -0.8  & -0.8  & -5.5  & -42.0 & -40.4 & -121.8 & -57.6 & -106.9 & -529.2 \\
& Second-stage score (€) & 3961.4 & 2033.3 & 670.9 & 182.2 & 749.6 & 174.5 & 567.6 & 897.0 & 534.4  & 594.6 & 2002.5 & 3656.5 \\
& Total score (€)        & 3779.8 & 2017.8 & 669.0 & 181.4 & 748.8 & 169.0 & 525.6 & 856.5 & 412.6  & 537.0 & 1895.6 & 3127.3 \\
& Gap to S-a (€)  & 1598.4 & 630.8  & 1664.9  & 53.9  & 123.4  & 21.7  & 85.1 & 341.2 & 243.3 & 339.4 & 1060.1 & 2365.6 \\
& Gap to S-a (\%)      & 73.3\% & 45.5\% & 33.1\% & 42.2\% & 19.7\% & 14.7\% & 19.3\% & 66.2\% & 143.7\%  & 171.9\% & 127.1\% & 310.6\% \\ \hline
\multirow{5}{*}{All-arbitrage}                         & First-stage score (€) & -169.9 & -14.3  & -1.6  & -0.9  & -0.8  & -5.1  & -41.6 & -42.2 & -121.3 & -56.2 & -93.1 & -485.7 \\
& Second-stage score (€) & 3852.3 & 2220.9 & 752.5 & 236.0 & 864.2 & 201.6 & 623.6 & 1047.9 & 606.7  & 628.2 & 2036.8 & 3961.5 \\
& Total score (€)        & 3682.4 & 2206.6 & 750.9 & 235.1 & 863.4 & 196.5 & 581.9 & 1005.7 & 485.3  & 572.0 & 1943.6 & 3475.8 \\
& Gap to S-a (€)  & 1501.0 & 819.6  & 248.4  & 107.5  & 238.0  & 49.1  & 141.3 & 490.4 & 316.0 & 374.6 & 1108.9 & 2714.2 \\
& Gap to S-a (\%)      & 68.8\% & 59.1\% & 49.4\% & 84.2\% & 38.1\% & 33.3\% & 32.1\% & 95.1\% & 186.6\%  & 189.7\% & 132.8\% & 356.3\% \\
\hline
    \end{tabular}
    }
\end{center}
\end{table*}

Three scenarios are examined: In the first scenario, all aggregators are stand-alone aggregators using the MADDPG algorithm with two siloed sub-agents (S-a). In the second scenario, only aggregator 5 on bus 5 is designated as the arbitrage aggregator with collaborative sub-agents, while the remaining are stand-alone aggregators (one-arb). Aggregator 5 was specifically chosen for this role due to its high generation capacity and significant market share, attributes that make it well-suited to explore the potential for market power abuse. In the third scenario, all aggregators participate in the market as arbitrage aggregators with collaborative sub-agents (all-arb).\looseness=-1

Figure \ref{fig:converge_agent} illustrates the convergence of ten aggregators using HMARL and two independent MARL algorithms across three scenarios—S-a, one-arb, and all-arb—during the evaluation week, represented by the green, blue, and red lines, respectively. The y-axis represents the aggregator score, which indicates the cumulative rewards earned in the market during the evaluation week, characterized by reduced net costs in euros. The x-axis denotes the number of training episodes for the aggregators, ranging from 0 to 5000. It can be seen that all the aggregators converge to a stable trading policy in all scenarios at the end of the training episode. The aggregators with the arbitrage strategy achieve a higher profit than the stand-alone strategy, reflected by the lowest scores. In the one-arb scenario, the performance of aggregator 5 with the arbitrage strategy outperforms the stand-alone strategy based on two independent MADDPG algorithms, which suggests that the proposed HMADDPG algorithm with the arbitrage strategy solves the proposed two-stage Markov game by offering superior performance and leading to higher cumulative rewards. In the all-arb scenario, nearly all aggregators achieve higher scores than in the S-a scenario. However, profitability varies. Some aggregators see significant score improvements, while others experience smaller gains. This difference arises because when all aggregators use an arbitrage strategy. Unlike the one-arb scenario, where a single arbitrage strategy competes against stand-alone strategies to demonstrate superior performance, the all-arb scenario becomes an equilibrium-seeking problem. In this context, each aggregator's score may or may not lead to a better outcome. A detailed market analysis for both scenarios is presented in the following subsections.  \looseness=-1

\subsection{Market Result Analysis: One-arbitrage}

\begin{figure*}
    \centering
  \subfloat[LEM net cost.\label{fig:LEM_cost}]{%
       \hspace{-0.1cm}\includegraphics[scale=0.46]{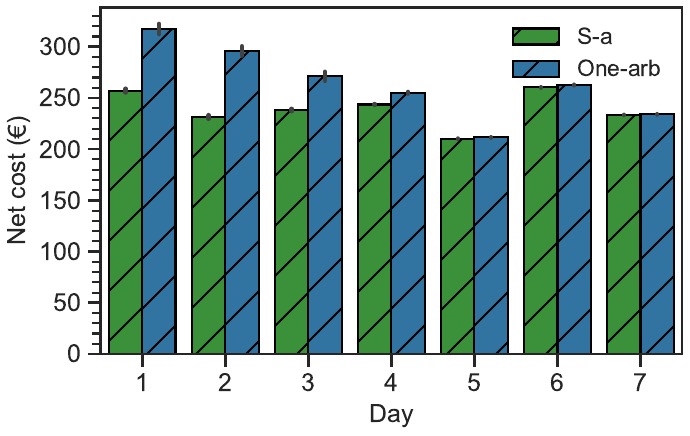}}
    \hfill
  \subfloat[LFM net cost.\label{fig:LFM_cost}]{%
        \includegraphics[scale=0.46]{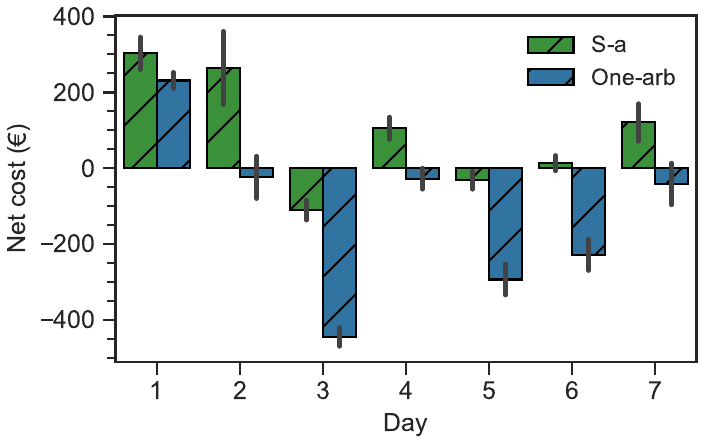}}
    \hfill
  \subfloat[Balancing market net cost.\label{fig:bal_cost}]{%
        \includegraphics[scale=0.46]{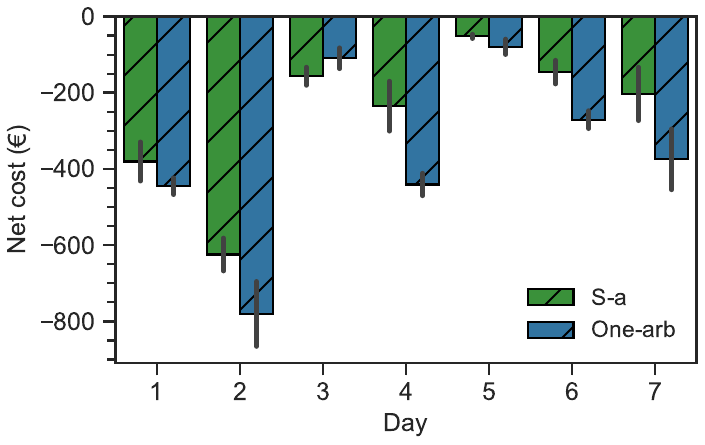}}\hspace{0.6cm}
  \caption{Market result of one test week}
  \label{fig:market_result} 
\end{figure*}

This subsection provides a detailed analysis of the one-arb scenario where aggregator 5 is designed as the arbitrage aggregator, while the others are stand-alone aggregators. Figure \ref{fig:score_box} and \mbox{Table \ref{tab:profit}} present detailed market results for aggregator 5 under both arbitrage and stand-alone bidding strategies within the two-stage Markov game on 12 test weeks. In Figure \ref{fig:score_box}, the x-axis represents the indices of the test weeks, and the y-axis represents the aggregator scores across multiple random seeds for each test week, visualized in a box plot. It is evident that the proposed HMADDPG algorithm with an arbitrage strategy consistently achieves a higher average score across all 12 test weeks compared to the stand-alone strategy, with performance gaps ranging from €21.7 ($14.7\%$) to €2365.6 ($310.6\%$), and an average score gap of €690.0 ($95.5\%$).   \looseness=-1

\begin{figure}[bt]
\centering
    \hspace{-0.0cm}
    \includegraphics[scale=0.48]{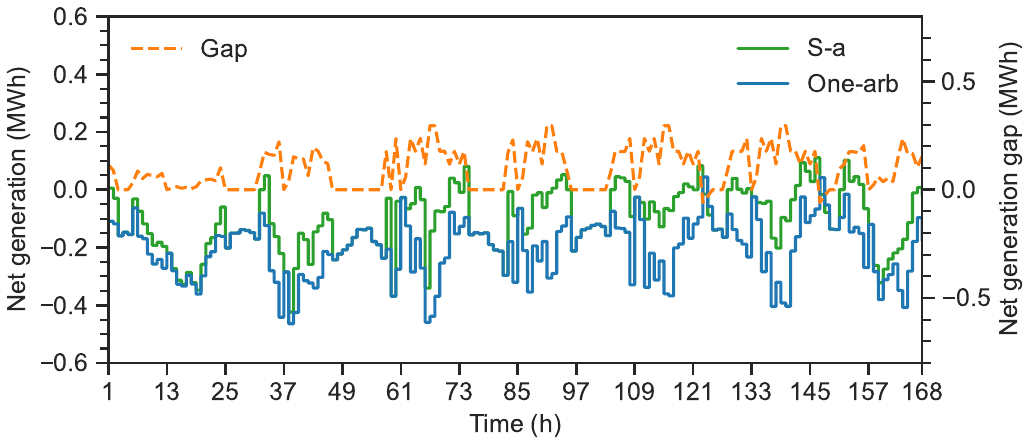}
\caption{LEM result in one test week.}
\label{fig:LEM_net_gen}
\end{figure}

%\begin{figure}[bt]
%\centering
%    \hspace{-0.0cm}
%    \includegraphics[scale=0.48]{figure/flexibility_quantity_price.pdf}
%\caption{LFM result in one test week.}
%\label{fig:LFM_test_week}
%\end{figure}

\begin{figure}
    \centering
  \subfloat[LFM quantity.\label{fig:flex_quantity}]{%
       \includegraphics[scale=0.48]{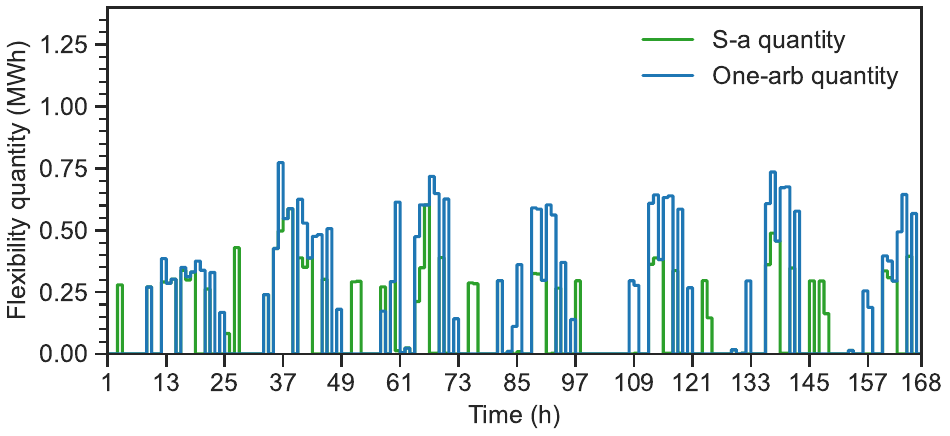}}
    \hfill
  \subfloat[LFM price.\label{fig:flex_price}]{%
        \includegraphics[scale=0.48]{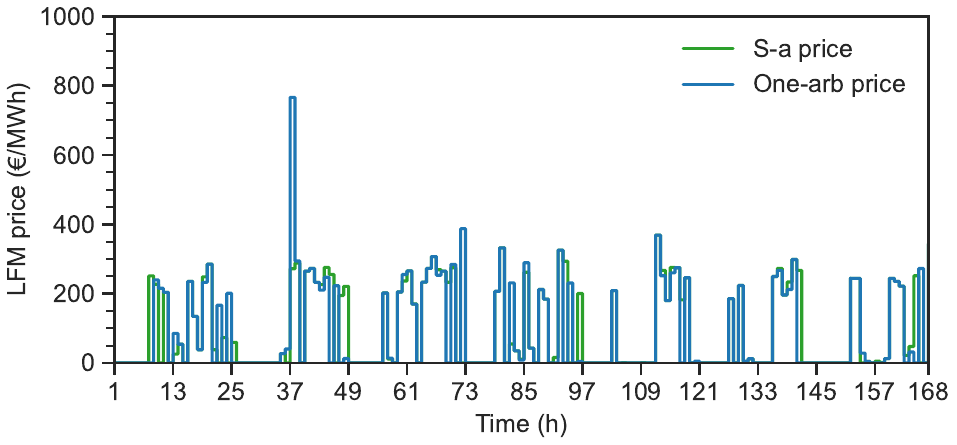}}
  \caption{LFM result in one test week.}
  \label{fig:LFM_test_week} 
\end{figure}

\begin{figure}
    \centering
  \subfloat[Positive imbalance.\label{fig:BAL_pos}]{%
       \includegraphics[scale=0.48]{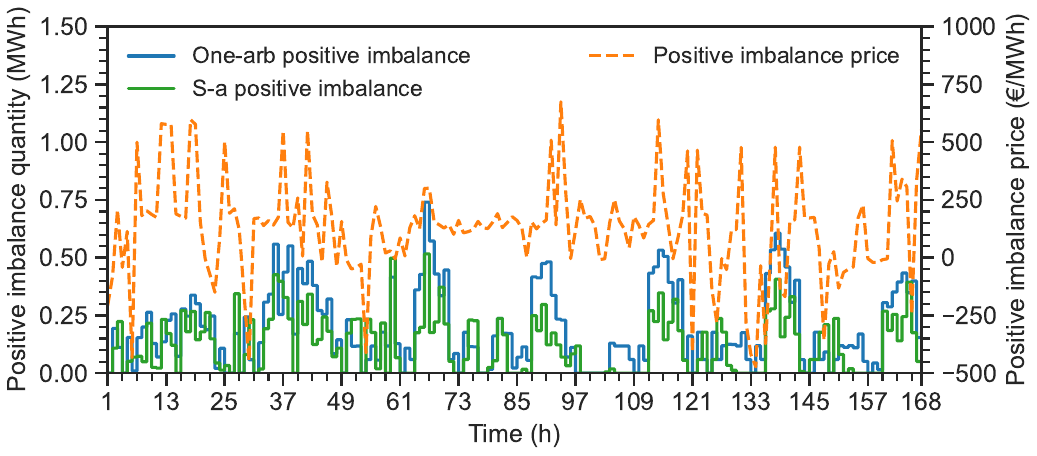}}
    \hfill
  \subfloat[Negative imbalance.\label{fig:BAL_neg}]{%
        \includegraphics[scale=0.48]{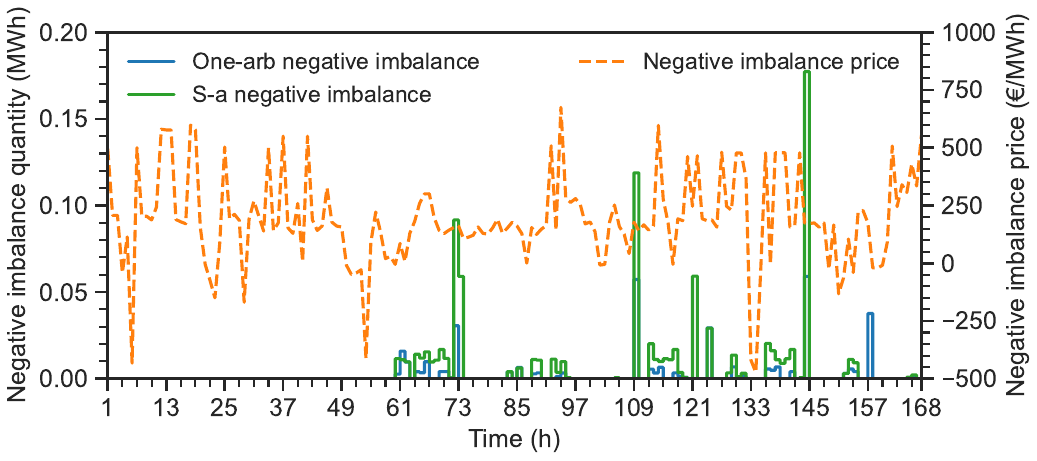}}
  \caption{Balancing market results for aggregator 5}
  \label{fig:BAL_result} 
\end{figure}

Table \ref{tab:profit} presents a detailed comparison of the performance of the two strategies for aggregator 5 at each stage. In the first stage of the two-stage Markov game, the score reflects the net cost in the LEM. In the second stage, the score accounts for income and operational costs from providing flexibility services in the LFM, as well as balancing market payments resulting from discrepancies between the actual post-LFM dispatch and the LEM-scheduled profiles due to deviations in flexibility service power profiles. If the deviation is positive (i.e., generation exceeds scheduled profiles), the aggregator receives a payment from the balancing market at positive imbalance prices. Conversely, if the deviation is negative (i.e., generation falls short of scheduled profiles), the aggregator incurs a payment to the balancing market at negative imbalance prices. The results show that the stand-alone aggregator, with two siloed sub-agents, achieves high scores with low net costs in the first stage, consistently approaching €0 across all test weeks. This is because the primary sub-agent operates independently of the secondary sub-agent, aiming to minimize net costs in the LEM by generating as much energy as possible with available cheap energy resources, without any physical quantity withholding. As a result, the net cost equals the reference cost $\bar{c}^{LEM}$ in (\ref{eq:reward1}), causing the first-stage score to approach €0. In contrast, the arbitrage aggregator with collaborative sub-agents exhibits lower first-stage scores across all test weeks, averaging €-92.0. This strategy of withholding generation from energy resources and importing from the external market at higher hourly prices leads to increased net costs in the first stage. However, the collaboration among the sub-agents of the arbitrage aggregator, based on the proposed HMADDPG framework, enables the primary sub-agent to make decisions by considering the overall score across all stages. Consequently, the secondary sub-agent of the arbitrage aggregator achieves a high score in the second stage, with an average value of €1335.4, which is $101.6\%$ higher than that of the stand-alone strategy (€662.3), resulting in a higher total score. \looseness=-1

The results demonstrate that the proposed HMADDPG algorithm enables the aggregator to achieve higher scores by facilitating collaboration among sub-agents. In contrast to the stand-alone strategy, where siloed sub-agents use two independent MADDPG algorithms that focus solely on minimizing costs at each stage, the arbitrage strategy allows for short-term cost increases in the first stage. This approach leads to substantial reductions in second-stage expenses, ultimately improving the overall score. \looseness=-1

To have a detailed analysis of the market results for a single test week, Figure \ref{fig:market_result} illustrates the net costs incurred by aggregator 5 for each day of the test week under both bidding strategies. Consistent with the discussion in the previous section, the aggregator achieves lower net costs under the HMARL-based arbitrage strategy in both the LFM and the balancing market than the stand-alone strategy using two independent MADDPG algorithms for all 7 days over the week by sacrificing short-term gains in the LEM. Figure \ref{fig:LEM_net_gen} illustrates the net generation of aggregator 5 during the test week. Figure \ref{fig:LFM_test_week} displays the price and quantity of flexibility services provided by aggregator 5 within the LFM during the same period. It can be seen that aggregator 5's arbitrage strategy involves periodic withholding of generation capacity at specific time slots, leading to net generation gaps from the stand-alone strategy. This behavior can contribute to grid congestion or voltage issues within the grid. Consequently, the DSO may need to procure a significant quantity of flexibility services from aggregator 5 in the LFM to mitigate these issues and maintain the safe and stable operation of the DN. More specifically, during the test week, aggregator 5 provided 29.9 MWh of flexibility service quantities in total, which is $68.5\%$ higher than the stand-alone strategy (17.8 MWh). Simultaneously, aggregator 5, under the arbitrage strategy, manipulated flexibility prices at specific time steps without significantly increasing the overall price of flexibility services compared to the stand-alone strategy. For example, at time step 37, the flexibility price peaked at €766.3/MWh under the arbitrage strategy, which is $107.7\%$ higher than the maximum flexibility price of €369.0/MWh under the stand-alone strategy at time step 112. At the same time, the average flexibility price of aggregator 5 under the arbitrage strategy is €90.7/MWh, nearly identical to the average flexibility price of €91.4/MWh under the stand-alone strategy. Consequently, it can be concluded that the aggregator achieves a lower net cost under the arbitrage strategy in the LFM by increasing the DSO’s demand for flexibility services and raising flexibility prices at specific time steps when selling large quantities of flexibility services. \looseness=-1

Figure \ref{fig:BAL_result} illustrates the balancing market results for aggregator 5, highlighting the performance in handling positive and negative imbalances under both arbitrage and stand-alone strategies. Figure \ref{fig:BAL_pos} presents the results for positive imbalance shows that the arbitrage strategy yields higher quantities of positive imbalance compared to the stand-alone strategy. This effect is especially significant during periods of high positive imbalance prices, as depicted by the dashed blue line. It indicates that the arbitrage strategy effectively exploits favorable positive imbalance prices to increase positive imbalance payments. Figure \ref{fig:BAL_neg} shows that negative imbalance quantities are lower with the arbitrage strategy than with the stand-alone approach. Notably, these negative imbalances primarily occur during hours of relatively low negative imbalance prices, indicating that the arbitrage strategy not only optimizes positive imbalance generation but also strategically minimizes exposure to negative imbalances. In summary, the results confirm that the proposed HMADDPG algorithms for the arbitrage strategy enhance the aggregator's performance by increasing positive imbalances during high-price periods and decreasing negative imbalances. This approach improves the overall net cost reduction in the balancing market through effective leveraging of price signals.\looseness=-1

\subsection{Market Result Analysis: All-arbitrage}

\begin{figure}[bt]
\centering
    \hspace{-0.0cm}
    \includegraphics[scale=0.48]{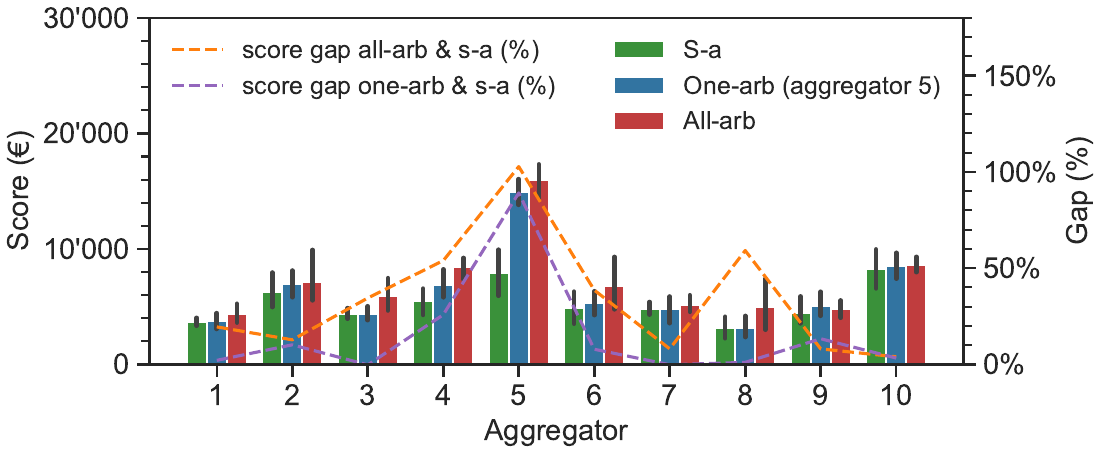}
\caption{Score of the aggregators.}
\label{fig:score_all_arb}
\end{figure}

In this section, the market outcomes of the all-arb scenario, where all participants use the HMARL algorithm to arbitrage across local markets, are analyzed. Figure \ref{fig:converge_agent} shows that the training process of the aggregators converges after 5000 episodes, indicated by the red line. Figure \ref{fig:score_all_arb} then presents the market outcomes for the aggregators during the test weeks. The x-axis represents the indices of the aggregators, while the y-axis shows the cumulative score of the aggregators in the test weeks, reflecting the profit they earned from the local markets. The red bars represent the total scores of aggregators using the all-arbitrage strategy, while the blue and green bars correspond to the one-arbitrage and stand-alone strategies, respectively. The orange and purple dashed lines illustrate the percentage score gaps between the all-arbitrage and stand-alone scenarios, and the one-arbitrage and stand-alone scenarios, respectively. 

The results show that all aggregators in the all-arb scenario outperform those in the S-a scenario in terms of score, indicating that the arbitrage strategy using HMARL with collaborative sub-agents leads to higher profits for all participants. Compared to the S-a scenario, aggregator 5 exhibits the most significant score increase of approximately $120\%$, which may be attributed to its substantial market share, enhanced flexibility capacities, and potentially strategic location at a critical system node, thereby leading to a higher profit. Besides that, aggregator 8 also achieves a significant score increase of $70.6\%$. Compared to the one-arb scenario, aggregators in the all-arb scenario also obtain higher profits, though the extent of the increase varies. Notably, aggregators 5 and 8 experience above-average profit gains of more than $80\%$, while the others see more modest or minimal increases (aggregators 9 and 10). These disparities are because of the market equilibrium effects when all aggregators adopt the arbitrating strategy. In the all-arb scenario, where each aggregator withholds capacity in the LEM and LFM, competition intensifies compared to the one-arb scenario, where only one aggregator arbitrages across local markets. The average score gap across all aggregators is $38.9\%$. \looseness=-1

\begin{figure}[bt]
\centering
    \hspace{-0.0cm}
    \includegraphics[scale=0.48]{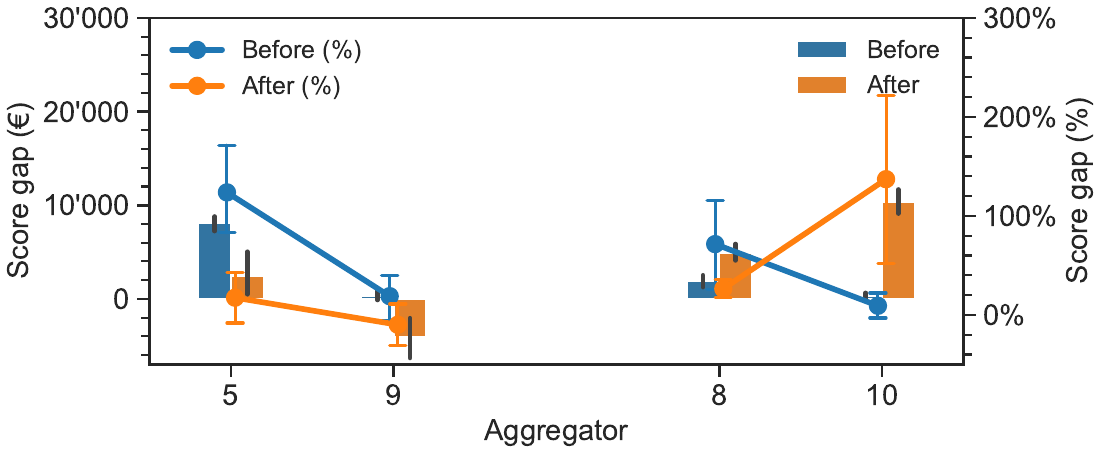}
\caption{Sensitivity analysis of the aggregators.}
\label{fig:score_sensitivity}
\end{figure}

\begin{table}[tb]
\caption{Total profit of the aggregators}
\label{tab:total profit}
\small
\begin{center}
\renewcommand\arraystretch{1.2}
\setlength{\tabcolsep}{2.5mm}{
\begin{tabular}{ccc}
\hline 
        & Evaluation week & Test week \\ \hline
Stand-alone/S-a (€) & 7238.5 & 5054.7\\
One-arbitrage/One-arb (€)    & 9644.9 & 6120.0 \\
All-arbitrage/All-arb (€)    & 12'986.0  & 7108.1\\
 \hline
\end{tabular}
}
\end{center}
\end{table}

To examine the profit disparities among aggregators in more detail, a sensitivity analysis is conducted to assess the impact of network location on aggregator performance by switching the network locations of aggregators 5 and 9, and aggregators 8 and 10. Figure \ref{fig:score_sensitivity} illustrates the results of this analysis. It consists of bar plots comparing the score (profit) gap between arbitrage and stand-alone strategies for four aggregators before and after switching their network locations. The x-axis lists the aggregator indices, and the left-hand y-axis represents the score gap in euros. Blue bars indicate the score gap before the location switch, while orange bars show the score gap after switching the locations of aggregators. Additionally, line plots represent the score gap between arbitrage and stand-alone strategies in percentage, providing a relative measure of performance changes. The analysis shows clear changes in market power. Aggregator 5's score gap falls from €10000 to €3000, and the line plot underscores this steep decline in percentage from $142.8\%$ to $18.6\%$. Aggregator 9 obtains a €300 loss after the switch, which also results in a lower profit after switching locations. Switching locations impacts aggregators 8 and 10 differently. Aggregator 8's absolute score rises from €2531.8 to €6171.1, but its arbitrage strategy’s relative advantage over the stand-alone strategy drops from $81.1\%$ to $27.0\%$, reducing the strategy's benefit. Conversely, aggregator 10 sees substantial gains: its score jumps from €462.8 to €12'593.1, and its relative advantage increases from $8.3\%$ to $130.6\%$. These findings indicate that network location, especially at critical network nodes, significantly impacts market power in the LFM, resulting in diverse profit outcomes driven by market equilibrium dynamics. Consequently, the resulting equilibrium unevenly redistributes surplus, with some aggregators securing larger profits while others face heightened competitive pressure that limits their gains. \looseness=-1

Table \ref{tab:total profit} presents the average total profits of all aggregators for the evaluation and test weeks. In both periods, the S-a scenario yields the lowest profits, followed by the one-arb scenario, while the all-arb scenario results in the highest profits. Compared to the S-a scenario, the all-arb scenario achieves €5747.5 ($79.4\%$) and 2053.4 ($40.6\%$) higher profits in the evaluation and test weeks, respectively. Relative to the one-arb scenario, the all-arb scenario delivers 3341.1 ($34.6\%$) and 988.1 ($16.1\%$) higher profits in these respective periods. Thus, full arbitrage adoption maximizes collective surplus but results in varied outcomes driven by equilibrium dynamics. \looseness=-1

\section{Conclusion and Outlook}\label{sec:conclusion}

This paper proposes an HMARL method with two sub-agents, enabling the aggregator to engage in arbitrage across local markets, comprising an LEM for trading energy, an LFM for flexibility service trading, and a balancing market for managing the aggregators' deviations from their scheduled profiles. The bidding strategies of aggregators in the local markets are formulated as a two-stage Markov game: the first stage includes the LEM, while the second stage encompasses both the LFM and the balancing market. In this model, aggregators are self-interested, non-cooperative, and each comprises two sub-agents. Without the arbitrage strategy, these sub-agents function in silos with two independent MADDPG algorithms: the primary sub-agent focuses solely on first-stage profits, and the secondary sub-agent concentrates on second-stage profits, with no communication between them. Conversely, when employing the arbitrage strategy with the HMADDPG algorithm, the sub-agents collaborate to minimize overall net costs across the local markets. The case study demonstrates that the aggregator employing the proposed HMADDPG algorithm sacrifices short-term gains by incurring higher first-stage net costs in the LEM to achieve substantial reductions in second-stage expenses within the LFM and balancing market. For the test weeks, the arbitrage strategy resulted in an average total profit increase of €2053.3 (40.6\%), achieved through substantial savings in the LFM and the balancing market, despite incurring higher initial costs in the LEM. Future research will concentrate on detecting and averting market power behaviors within local markets. The proposed HMARL scheme also requires benchmarking against existing rule-based or optimization-based strategies to assess its real-world applicability. Additionally, analyzing the collusion of aggregators will provide further insights into market dynamics and potential for abuse.

\looseness=-1

\section*{Acknowledgement}

This work was supported by NCCR Automation, grant agreement 51NF40\_180545 from the Swiss National Science Foundation. This publication is also part of the research program ‘MEGAMIND – Enabling distributed operation of energy infrastructures through Measuring, Gathering, Mining and Integrating grid-edge Data', (partly) financed by the Dutch Research Council (NWO), through the Perspectief funding instrument under number P19-25.

%\bibliographystyle{cas-model2-names}
\bibliographystyle{IEEEtran}

\vspace{-0.5em}
%\bibliography{cas-refs}

\end{document}